# Improvement and Stabilization of Output Voltages in a Vertical Tidal Turbine Using Intelligent Control Strategies

F. Philibert ANDRINIRINIAIMALAZA
Laboratoire du Génie Electronique Informatique
Institut Supérieur des Sciences et Technologies de Mahajanga
University of Mahajanga
Mahajanga, Madagascar
philibert.andriniriniaimalaza@gmail.com

Nour Mohammad MURAD
Laboratoire de Physique et Ingénierie Mathématique pour l'Energie L'Environnement et le Bâtiment (PIMENTLab)
IUT, Département Réseaux et Télécoms, Université de La Réunion
Saint Pierre, La Reunion
nour.murad@univ-reunion.fr

Telesphore RANDRIAMAITSO
Laboratoire du Génie Electrique
Institut Supérieur des Sciences et Technologies de Mahajanga
University of Mahajanga
Mahajanga, Madagascar
rmaitsotelesphore@univ-reunion.fr

Habachi BILAL
Laboratoire des Technologies Innovantes (LTI),
Université de Picardie Jules Verne
80000 Amiens, France
habachibilal7@gmail.com

Nirilalaina RANDRIATEFISON
Ecole Normale Supérieure - Université of Antananarivo,
Antananarivo, Madagascar
randriatefison@yahoo.fr

Ruffin MANASINA
Laboratoire du Génie Electrique
Institut Supérieur des Sciences et Technologies de Mahajanga
University of Mahajanga
Mahajanga, Madagascar
ruffin.manasina@gmail.com

Charles Bernard ANDRIANIRINA
Laboratoire du Génie Electronique Informatique
Institut Supérieur des Sciences et Technologies de Mahajanga
University of Mahajanga
Mahajanga, Madagascar
nirina.cha.ca@gmail.com

Blaise RAVELO
Nanjing University of Information Science & Technology (NUIST), School of Electronic & Information Engineering, Nanjing, Jiangsu, China
blaise.ravelo@nuist.edu.cn

*Abstract*—This study investigates the improvement and stabilization of alternating current (AC) and direct current (DC) output voltages in a Permanent Magnet Synchronous Generator (PMSG) driven by a vertical-axis tidal turbine using advanced control strategies. The research integrates artificial intelligence-based techniques to enhance voltage stability and efficiency. Initially, the Maximum Power Point Tracking (MPPT) approach based on Tip Speed Ratio (TSR) and Artificial Neural Network-Fuzzy Logic (ANN-Fuzzy) controllers is explored. To further optimize performance, Particle Swarm Optimization (PSO) and a hybrid Artificial Neural Network-PSO (ANN-PSO) methodology are implemented. These strategies aim to refine the reference rotational speed of the turbine while minimizing deviations from optimal power extraction conditions. The simulation results for a tidal turbine operating at a water flow velocity of 1.5 m/s demonstrate that the PSO-based control approach significantly enhances voltage stability compared to conventional MPPT-TSR and ANN-Fuzzy controllers. The hybrid ANN-PSO technique further improves voltage regulation by dynamically adapting to system variations and providing real-time reference speed adjustments. This research highlights the potential of AI-based hybrid optimization in stabilizing the output voltage of tidal energy systems, thereby increasing reliability and efficiency in renewable energy applications.

*Keywords—vertical axis tidal turbines, permanent magnet synchronous generator, maximum power point tracking, renewable energy systems, voltage Stabilization, Artificial Intelligence Applications, Hybrid Optimization*

## I. Introduction

An alternative approach to harnessing tidal energy is the development of tidal stream devices, which represent a more efficient method of power generation compared to tidal barrage plants. This technology is increasingly attractive for power generation, especially in isolated areas or developing countries, due to environmental risks reductions and fewer capacity limitations. Research and development efforts on this emerging energy source are rapidly advancing, with engineers focused on creating environmentally safe turbines and marine scientists studying their potential ecological impact [1].

Hydropower, specifically tidal power, has emerged as one of the most promising renewable energy sources. Tidal turbines, similar to wind turbines, capture the kinetic energy from river or ocean currents, Figure 1-a, atypically requiring an average water current speed of more than 1 m/s to generate power [2].

Although horizontal-axis tidal turbines are known for their higher efficiency and have been adopted by numerous renewable energy companies worldwide, the vertical-axis tidal turbines present different characteristics and offer some distinct advantages, such as being cheaper to produce, easier to install due to their ability to place the generator on top, and better suited for floating systems [3].

Tidal turbines typically use permanent magnet synchronous generators (PMSG) due to their excellent mechanical-to-electrical conversion efficiency, which can reach close to 99%, outperforming asynchronous machines. Additionally, PMSGs can be coupled or uncoupled with a speed multiplier [4]. Despite the advantages, studies have shown that vertical-axis turbines generate pulsating torque causing instability in the voltage or power output, as seen in the voltage oscillations linked to mechanical vibrations induced by the water flow [5].

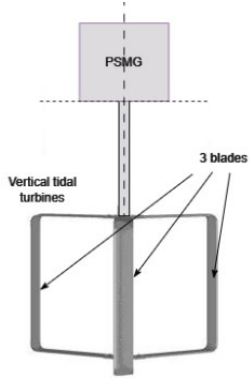

Fig. 1. Vertical axis turbine arrangements

To address this challenge, control systems are often employed. Several control methods have been proposed, including PI controllers [6], PID controllers [7], fuzzy logic [8], genetic algorithms [9] and artificial neural networks (ANN) [10].

In this study, we propose the use of diverse algorithm such an ANN-fuzzy strategy to control the tidal turbine system. ANN-fuzzy systems are particularly effective in managing imprecise or incomplete data while recognizing complex patterns [11]. Furthermore, particle swarm optimization (PSO) and artificial neural networks are integrated into this control strategy to optimize the system performance. PSO, an efficient optimization technique inspired by natural swarm behaviors, is used to fine-tune the controller parameters, improving the overall system's energy output [12].

The integration of ANN with PSO (ANN-PSO) enhances the optimization process, allowing the controller to adapt dynamically to the varying conditions of the tidal flow, thus maximizing power extraction [13].

This study focuses on the performance of a tidal turbine with a vertical axis of rotation coupled to a direct-drive PMSG. The primary objective is to analyze the impact of the pulsating torque on the output voltage of the PMSG and compare the effectiveness of energy optimization strategies. We aim to assess the system's regulation (%), efficiency (%), response time (s), performance, power factor, and stability, using both the MPPT TSR, ANN-fuzzy, PSO, and ANN-PSO hybrid optimization approach.

This article is organized into five sections. In section 2, the system under study is defined and is described as well as its architecture. Then, the MPPT controllers are explained after, for the TSR, ANN-fuzzy, PSO, and ANN-PSO. Simulation results are presented in section 3 and followed by discussions to assess the contribution in section 4. Finally, section 5 presents the conclusions derived from the preceding sections.

## II. MATERIAL AND METHOD

### A. The proposed model and its parameters

Tidal turbines, similar to wind turbines, extract kinetic energy from river or sea currents, requiring a minimum current speed of 1 m/s to generate power [14].

The use of a PMSG is preferred due to its high performance enabling direct drive operation without a speed multiplier, unlike other generators such as asynchronous generator cage (AGC), synchronous generator (SG), … [15].

The system under study is shown in Figure 2. It is composed by tidal turbine model, MPPT controller, PMSG with its vector control, and pulse width modulation (PWM) signal generation for the rectifier's power transistors [16].

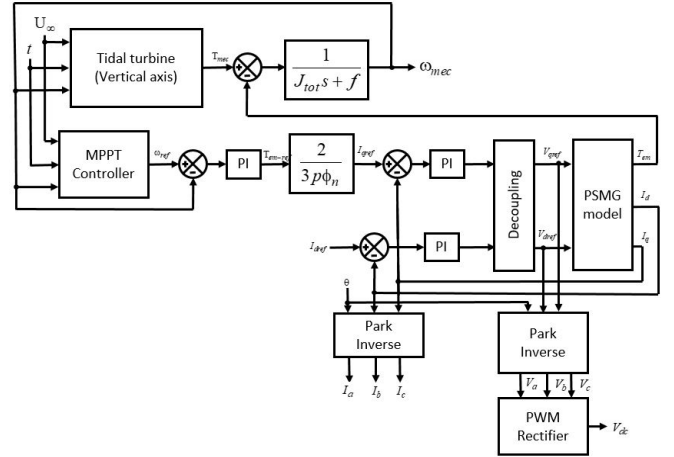

Fig. 2. The system with rectifier and its controller unit

The tidal turbine utilizes MPPT strategies, respectively, based on an optimum tip-speed ratio (TSR) [17], an ANN-fuzzy approach [18], a PSO algorithm [19], and an ANN-PSO strategy [20]. Additionally, the PMSG is using the Park model and controlled by vector control with a PI corrector [21]. The PWM rectifier is based on Insulated Gate Bipolar Transistors (IGBTs).

Mechanical power Pm generated by the turbine is a fraction $C_P$ of the fluid's hydrodynamic power [22], expressed by:

$$P_m = \frac{1}{2} C_p \rho A U_\infty^3 \quad (1)$$

where $\rho$ is water density (kg/m³), A is the rotor's cross-sectional area (m²), and $U_\infty$ is the water current velocity (m/s).

The advance ratio $\lambda$ is a key factor in turbine performance [23], defined as:

$$\lambda = \frac{\omega r}{U_\infty} \quad (2)$$

where $\omega r$ is the peripheral velocity of the blades (m/s).

Based on equations (1) and (2) and by analyzing the relationship between the power coefficient $C_P$ and the tip-speed ratio $\lambda$, a double-multiple stream-tube theory (DMST) method is used [24]. Table I outlines the physical parameters of the tidal turbine.

TABLE I. PHYSICAL PARAMETERS OF THE TIDAL TURBINE

| Parameters | Symbol | Value |
|---|---|---|
| Speed of sea currents | $U_\infty$ | 1.5 m/s |
| Rope length | $C$ | 156 mm |
| Turbine radius | $r$ | 455 mm |
| Turbine height | $h$ | 824 mm |
| Wedge angle | $\beta$ | 0 ° |
| Canopy inertia | $J_t$ | 1.5 kgm² |
| Friction | $f_t$ | 0.025 Nms/rad |
| Torque | $T_{mec}$ | 103.5 Nm (min) 197.9 Nm (max) |

| Advance ratio | λ | 0 to 4 |

To visualize and approximate the performance curve of the studied tidal turbine, from Table 1, a bell-shaped curve for CP(λ) is given in Figure 3. Maximum power coefficient (CP,max) and optimal tip-speed ratio (λopt) are critical parameters for optimizing turbine operation and validating numerical models (like DMST) against experimental data [23].

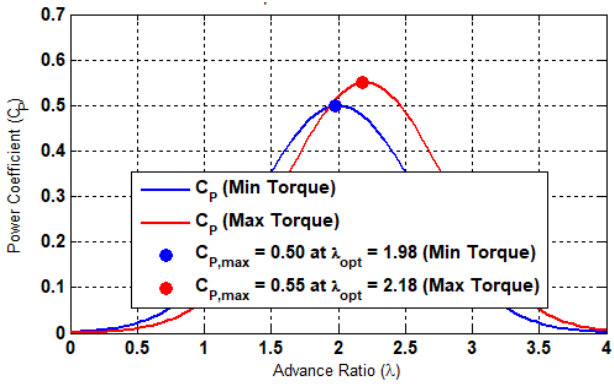

Fig. 3. Bell-shaped curve for $C_P(\lambda)$

For a maximum torque, the bell-shaped curve for $C_P(\lambda)$, Figure 3, is peaking at $\lambda = 2.18$ and $C_{P,max} = 0.55$.

With a comprehensive understanding of the system architecture and turbine performance outlined in this section, the focus now shifts to the design and implementation of maximum power point tracking (MPPT) controllers, which play a pivotal role in optimizing energy extraction from the tidal turbine under varying operational conditions.

*B. The MPPT controller methodologies*

To optimize power extraction and enhance the stability of tidal turbines under varying operational conditions, this section explores the design and implementation of four MPPT control strategies: MPPT-TSR, MPPT-ANN-Fuzzy, MPPT-PSO, and MPPT-ANN-PSO, each offering unique advantages in achieving maximum efficiency.

*1) The MPPT-TSR Controller*

A tidal turbine achieves its maximum power coefficient CP when operating at the optimal tip-speed ratio $\lambda_{opt}$. The Maximum Power Point Tracking (MPPT) control method ensures the turbine operates at $\lambda_{opt}$, thereby maximizing the mechanical power extracted from the water flow. This is achieved by continuously adjusting the turbine's rotational speed to match the optimal tip-speed ratio. Ensuring operation at λopt is critical, as deviations can lead to significant reductions in the power coefficient and, consequently, the turbine's efficiency [25].

As given in [23] and [24], the next step utilizes a proportional-integral (PI) regulator to maintain the optimal speed. The PI regulator compares $\omega_{ref}$, derived from the optimal TSR equation, with the measured mechanical speed ($\omega_{mec}$). The resulting error ($\varepsilon = \omega_{ref} - \omega_{mec}$) is used to adjust the torque reference, ensuring the turbine operates efficiently under varying flow conditions.

While the MPPT TSR controller effectively tracks the optimal tip-speed ratio, the MPPT ANN-Fuzzy controller introduces adaptive intelligence to further enhance power extraction under varying and complex operational conditions.

*2) The MPPT ANN-fuzzy Controller*

The proposed MPPT ANN-Fuzzy controller is designed to optimize power extraction from the tidal turbine under varying and dynamic operational conditions. By combining the adaptive capabilities of neural networks and the rule-based reasoning of fuzzy logic, this controller ensures precise and efficient adjustment of the turbine's mechanical speed set-point. This dual approach allows the system to respond effectively to extracted power and rotational speed changes.

The controller's conceptual framework is implemented in two key stages [26] [27]:
- First Stage: Artificial Neural Networks Modeling

A neural network is employed to model the turbine's mechanical power output as a nonlinear function of flow velocity ($U_\infty$) and rotational speed ($\omega_{mec}$). The neural network captures the turbine's complex dynamics, enabling accurate predictions of power generation under varying conditions.
- Second Stage: Fuzzy Logic for Variable Velocity

A fuzzy logic controller takes inputs such as changes in rotational speed ($\Delta\omega$), variations in power output ($\Delta P_{mec}(t)$), and time-based trends. The controller, based on predefined fuzzy rules, generates the reference rotational speed ($\omega_{ref}$), ensuring that the turbine operates near its maximum power point. This stage provides the flexibility to handle imprecise or incomplete input data, a common challenge in real-world turbine operations.

By integrating neural network-based modeling with fuzzy logic control, the ANN-Fuzzy controller achieves a balance between adaptability and robustness.

Building on the adaptability of ANN-Fuzzy controllers, further enhancements can be achieved by integrating advanced optimization techniques such as Particle Swarm Optimization (PSO), which provides a robust framework for handling the complexities of turbine dynamics.

*3) The MPPT-PSO controller*

Particle Swarm Optimization (PSO) is a population-based optimization algorithm inspired by the social behavior of birds flocking or fish schooling. It is widely used for solving optimization problems due to its simplicity and efficiency. PSO operates by iteratively improving candidate solutions, called "particles," within a search space, guided by their own experiences and the collective knowledge of the swarm [19].

Each particle in the swarm represents a potential solution to the problem. The movement of a particle is influenced by:
- Its personal best position ($p_{best}$), represents the best solution it has achieved.
- The global best position ($g_{best}$) represents the best solution the entire swarm finds.

Building on the MPPT-PSO approach, the MPPT-ANN-PSO controller incorporates neural networks to enhance adaptability and efficiency in maximizing power extraction.

*4) MPPT ANN-PSO controller*

Integrating ANNs into the PSO framework enhances the controller's capability by introducing a data-driven modeling approach.

The ANN learns the nonlinear relationship between the turbine's mechanical power, flow velocity, and rotational speed, in Figure 4. This predictive modeling provides a more precise and adaptive representation of the turbine's behavior,

serving as an informed foundation for the PSO optimization process [28] [29]:
- First stage: artificial neural network modeling

As outlined in subsection 3.2, the turbine's mechanical power is modeled using an artificial neural network, which accurately captures the nonlinear relationship between water flow velocity and rotational speed.
- Second stage: PSO for variable velocity

The PSO controller dynamically generates the reference rotational speed by adapting to changes in time, rotational speed, and power.

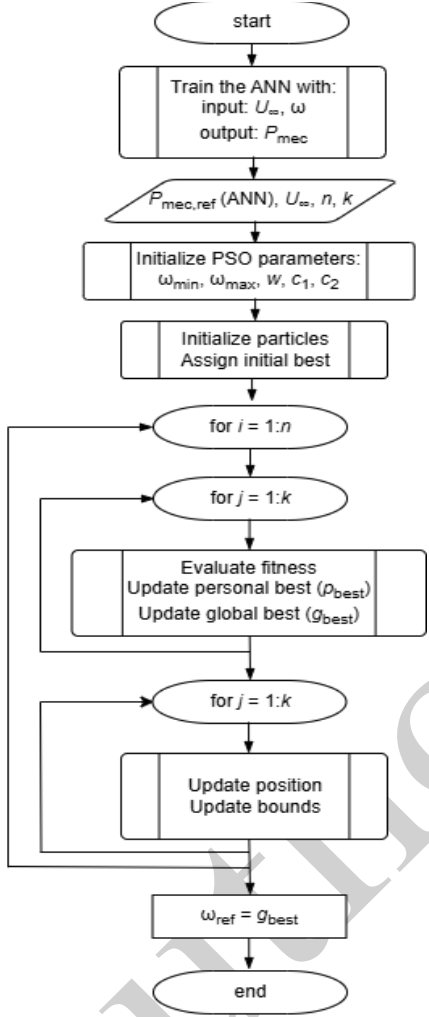

Fig. 4. MPPT ANN-PSO basic use

This mechanism ensures optimal power extraction, even in the presence of fluctuating flow conditions. By combining these two techniques, the ANN-PSO controller leverages the predictive power of ANNs and the optimization efficiency of PSO. This hybrid approach in Figure 4, enables the system to adjust to varying operational conditions dynamically, ensuring maximum power extraction with greater accuracy and responsiveness compared to using PSO alone

## III. RESULTS

The numerical simulation of the water turbine is conducted using the double multiple stream tube (DMST) method. In this model, the turbine rotor is divided into two halves: the upstream (front half-cycle) and downstream (rear half-cycle). The flow interacting with a rotor of radius $r$ is segmented into a series of adjacent stream tubes, each containing two sequential actuator plates [23] [24]. This approach captures the detailed interactions of the water flow with the turbine blades, allowing for accurate performance modeling.

Simulations are carried out in MATLAB/Simulink for the tidal turbine, PMSG, and their associated controllers. The physical parameters listed in Table 1 are used, with the turbine operating at an optimal advance ratio ($\lambda_{opt}$ = 2.18) that yields a maximum power coefficient ($C_{P,max}$ = 0.55) as shown in Figure 3.

The results presented below demonstrate the effectiveness of the model in capturing the turbine's behavior under varying conditions, highlighting its performance and stability.

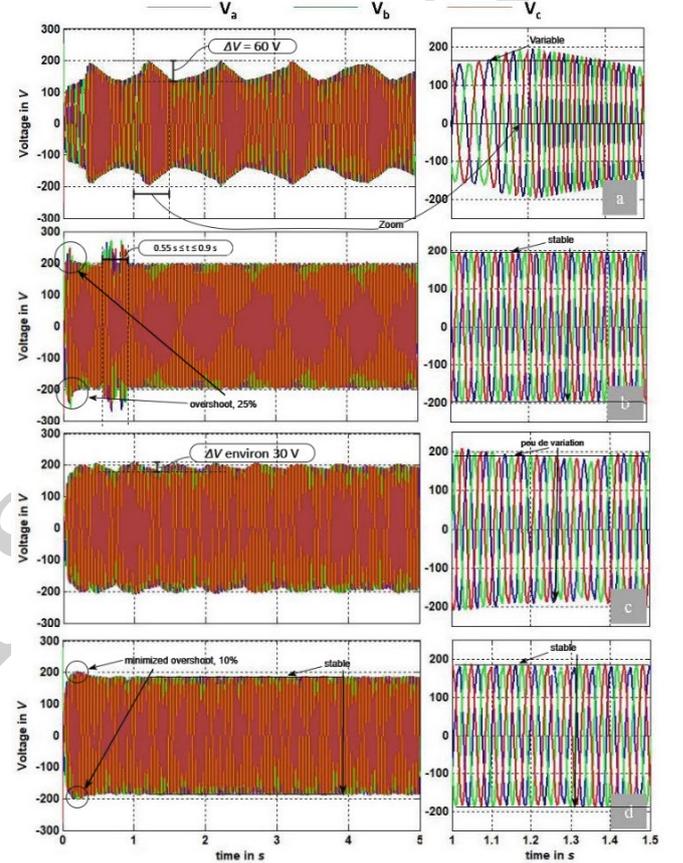

Fig. 5. Three phases output voltage diagram with MPPT a) TSR, b) ANN-fuzzy, c) PSO, and d) with ANN-PSO.

The curves in Figure 8 illustrate the evolution of the three-phase voltages ($V_a$, $V_b$, $V_c$) at the output of the PMSG under different regulation strategies, both in the absence and presence of regulation. The performance of the MPPT methods varies significantly, highlighting their unique characteristics and suitability for different operating conditions.

The MPPT-TSR controller exhibits a voltage range ($\Delta V_{AC} = V_{AC,max} - V_{AC,min}$) of 60 V, indicating significant variation in amplitude and relatively higher variability, which may impact overall system stability. This is attributed to its reliance on an average rotational speed as a reference, which does not fully account for the oscillating torque characteristic of vertical-axis turbines, resulting in more pronounced voltage fluctuations (Figure 5-a).

In contrast, the MPPT ANN-Fuzzy controller demonstrates superior performance by dynamically providing an instantaneous reference rotational speed that minimizes the error between measured and reference speeds, even under constant flow conditions. While it experiences an overshoot exceeding 25% and faces notable perturbations between 0.55 s and 0.9 s, it stabilizes effectively afterward, offering enhanced voltage stability (Figure 5-b).

The MPPT PSO controller eliminates overshoot, ensuring smoother operation with minimized voltage range ΔVAC close to 30 V and limited variability, making it suitable for maintaining consistent output under dynamic conditions. This is achieved by optimizing the reference rotational speed through a balance of global and local solutions, resulting in a smoother voltage profile (Figure 5-c).

The MPPT ANN-PSO controller combines the strengths of ANN and PSO to achieve superior performance, with an overshoot constrained to below 10%. By leveraging neural network-based torque modeling, it effectively predicts and adapts to oscillations, maintaining a stable envelope and ensuring robust operation even in fluctuating environments (Figure 5-d). These observations underline the trade-offs between stability, precision, and adaptability across the different MPPT strategies.

The DC output voltage over time is presented in Figure 6, highlighting the impact of torque oscillations on voltage stability.

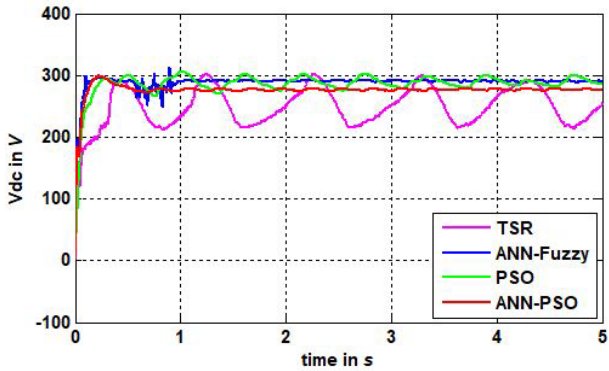

Fig. 6. DC output voltage with MPPT a) TSR, b) ANN-fuzzy, c) PSO, and d) with ANN-PSO.

Under an MPPT TSR, the oscillating torque effect is more pronounced in the system, leading to less noticeable voltage fluctuations. MPPT ANN-Fuzzy, PSO and ANN-PSO demonstrate that the control strategies effectively mitigate the torque oscillation effect, resulting in smoother and more stable DC voltage output.

## IV. Discussions

The effectiveness of the regulation strategies in optimizing the power dynamics and achieving steady-state conditions are illustrated. The transient phase duration, which characterizes the system's transition toward a stable state, decreases significantly across the different control strategies.

For the MPPT-TSR, the transient phase lasts approximately 1.5 seconds, while the MPPT ANN-fuzzy reduces this duration to 0.54 seconds. With the introduction of the PSO and ANN-PSO strategies, further improvements are observed: the transient phase is reduced to 0.35 seconds for the PSO and 0.25 seconds for the ANN-PSO. These results demonstrate the superior efficiency of the ANN-PSO approach in achieving rapid stabilization of the system, as shown in Table 2. This improvement underscores the ability of the regulation approaches to counteract the influence of pulsating torque and ensure consistent voltage stability during operation.

TABLE II. Transition phase duration

| | MPPT-TSR | MPPT ANN-fuzzy | MPPT PSO | MPPT ANN-PSO |
|---|---|---|---|---|
| Transition phase duration (s) | 1.5 | 0.54 | 0.35 | 0.25 |

On the performance side of the regulation systems, Table 3 summarizes the controller performance analysis between the three algorithms: MPPT TSR, MPPT ANN-fuzzy, MPPT PSO, and MPPT ANN-PSO.

TABLE III. Performance analysis between the MPPT algorithms: TSR, ANN-fuzzy, PSO, and ANN-PSO

| Considered parameters | MPPT TSR | MPPT ANN-fuzzy | MPPT PSO | MPPT ANN-PSO |
|---|---|---|---|---|
| $V_{DC,min}$ (V) $t \geq 1.0$ s | 214.09 | 284.08 | 270.09 | 273.00 |
| $V_{DC,max}$ (V) $t \geq 1.0$ s | 304.06 | 293.40 | 305.93 | 279.33 |
| $\Delta V_{DC}$ (V) | 89.97 | 9.33 | 35.85 | 6.33 |
| Response Time (s) | 1.5 | 0.54 | 0.35 | 0.25 |
| Voltage Regulation, $t \geq 0.25$ s (%) | 44.26 | 25.34 | 15.53 | 9.84 |
| Efficiency η (%) | 80 | 95 | 94 | 96 |
| HDR (%) | 20.02 | 5.0 | 5.1 | 1.75 |

The results clearly demonstrate the superiority of the MPPT ANN-fuzzy, MPPT PSO, and MPPT ANN-PSO controllers over the traditional MPPT-TSR in terms of both efficiency and performance. The MPPT ANN-fuzzy controller achieves a remarkable 95% efficiency and reduces harmonic distortion to just 5%, significantly outperforming the MPPT-TSR controller, which only reaches 80% efficiency and a higher HDR of 20%. Moreover, the optimization-based methods—MPPT PSO and MPPT ANN-PSO—offer notable improvements in both efficiency and harmonic distortion reduction. MPPT PSO achieves a 94% efficiency and an HDR of 5.1%, but the MPPT ANN-PSO method surpasses both, achieving the highest efficiency of 96% and the lowest HDR of 1.75%.

In addition to enhanced efficiency, the MPPT ANN-PSO controller demonstrates superior performance in several key areas. It exhibits the fastest response time, reaching steady-state conditions in just 0.25 seconds compared to the slightly longer times of MPPT PSO (0.35 seconds) and MPPT ANN-fuzzy (0.54 seconds). The MPPT ANN-PSO also provides better voltage regulation, ensuring more stable voltage output even under dynamic conditions.

## V. Conclusions

A vertical-axis water turbine with a radius of 455 mm and a height of 824 mm is considered in this study, operating at a minimum flow rate of 1.5 m/s. The turbine exhibits a minimum torque of 103.5 Nm and a maximum torque of 197.9 Nm. The impact of torque oscillations on the voltage output of the Permanent Magnet Synchronous Generator (PMSG) is analyzed by observing the trend of currents and voltages at the system's output. The results clearly demonstrate the superiority of the MPPT ANN-fuzzy, MPPT PSO, and MPPT ANN-PSO controllers over the traditional MPPT-TSR in terms of both efficiency and performance. The MPPT ANN-fuzzy controller, with a 95% efficiency and a reduced HDR of 5%, outperforms the MPPT-TSR controller, which only achieves 80% efficiency and a HDR of 20%.

The key distinction between MPPT PSO and MPPT ANN-PSO lies in their optimization mechanisms. While MPPT PSO uses Particle Swarm Optimization to search for the optimal parameters, MPPT ANN-PSO integrates an Artificial Neural Network (ANN) to enhance the search process, allowing for faster and more accurate convergence. Specifically, MPPT ANN-PSO achieves the quickest stabilization time, with a response time of 0.25 seconds, compared to 0.35 seconds for MPPT PSO. Additionally, it ensures more stable voltage output and better regulation over time.

The MPPT ANN-PSO optimization technique leads to a more precise tracking of the optimal operating point, resulting in superior performance in varying tidal conditions. This highlights the advantages of integrating artificial intelligence techniques, like ANNs, into traditional optimization algorithms like PSO. These findings confirm that MPPT ANN-PSO is the most effective method for enhancing both the efficiency and stability of tidal turbine systems.

Overall, these results emphasize the power of hybrid control strategies like MPPT ANN-fuzzy, MPPT PSO, and MPPT ANN-PSO in optimizing the performance of tidal energy systems. By combining optimization with neural network intelligence, these strategies provide adaptive and highly efficient solutions that ensure stable voltage output and minimal harmonic distortion, making them highly suitable for the dynamic nature of tidal energy generation.